\documentclass[aps,pre,twocolumn,showpacs,floatfix,groupedaddress]{revtex4}
\usepackage[dvips]{graphicx}
\begin{document}

\title{Two-dimensional melting far from equilibrium in a granular monolayer}

\author{J. S. Olafsen}
\email{jolafsen@ku.edu}
\affiliation{ Department of Physics and Astronomy, The University of Kansas,
Lawrence, KS 66045}

\author{J. S. Urbach}
\email{urbach@physics.georgetown.edu}
\affiliation{Department of Physics, Georgetown University, Washington, DC  20057}

\date{\today}

\begin{abstract}
We report an experimental investigation of the transition from a hexagonally ordered solid phase to a disordered liquid in a monolayer of vibrated spheres.  The transition occurs as the intensity of the vibration amplitude is increased.  Measurements of the density of dislocations and the positional and orientational correlation functions show evidence for a dislocation-mediated continuous transition from a solid phase with long-range order to a liquid with only short-range order.   The results show a strong similarity to simulations of melting of hard disks in equilibrium, despite the fact that the granular monolayer is far from equilibrium due to the effects of interparticle dissipation and the vibrational forcing.
\end{abstract}

\pacs{45.70.-n,05.70.Fh,05.70.Ln}

\maketitle

The nature of the melting transition of a two-dimensional solid has
been extensively studied since the pioneering work of Kosterlitz and
Thouless\cite{kosterlitz73} and subsequent extensions by Halperin,
Nelson, and Young\cite{young79,nelson79}, known collectively as the
KTHNY theory.  This continues to be an
active area of research \cite{dash99,moucka05}, and investigations of colloidal particles suggest that the specific form of interparticle interactions can play a role in determining the nature of the melting transition\cite{karn00}.
The particular system of hard disks has been extensively studied by computer simulation, and there is considerable
evidence that the melting transition is well
described by the KTHNY scenario (see e.g. \cite{watanabe05, binder02} and references therein).  That
scenario describes two-dimensional melting as consisting of two
continuous transitions, the first from a crystalline phase
characterized by algebraically decaying positional order and
long-range orientational order to a hexatic phase characterized by
short-range positional order and algebraically decaying orientational
order.  The hexatic phase melts into a liquid with only short-range
order.  Both transitions are mediated by topological defects, the
first by unbinding of dislocation pairs, the second by
unbinding of disclination pairs (see \cite{strandburg88} for a review).  

Single layers of identical macroscopic spheres subjected to mechanical
vibration undergo transitions that share, at least superficially, many
of the characteristics of equilibrium atomic and molecular melting
transitions\cite{pieranski78,clark90,olafsen98,strassburger00}. The
external vibration is necessary to keep the macroscopic spheres in
motion because of the inevitable loss of energy to friction and inelastic
collisions. As a result of the flow of energy into the system from the
external forcing and out of the system through dissipation, detailed
balance is violated, and the results of equilibrium statistical
mechanics are not generally applicable. We have shown directly that the
fluctuating velocities of the individual spheres\cite{olafsen98}
and their correlations\cite{prevost02} differ significantly from those
found in equilibrium systems.  Thus it may be expected that the
dynamics of the melting transition observed for driven, inelastic
spheres may be quite different from those observed in equilibrium.
However in at least one instance it has been shown that in certain
model systems the Gibbs distribution and detailed balance can be
recovered by coarse-graining over small length scales\cite{egolf00}.

In this paper we report an investigation of the melting transition of
a monolayer of identical spherical particles confined between two
vibrating plates.  As reported previously\cite{pieranski78,olafsen98}, 
a granular
monolayer at moderately high density and moderately low vibration
amplitude forms a ordered array, with each particle fluctuating about
sites arranged on a hexagonal lattice that extends across the entire
cell.  As the vibration amplitude is increased, the ordered lattice
`melts', and the spheres display disordered, fluctuating, liquid-like
dynamics. Here we show that this transition is continuous and is consistent with the KTHNY melting scenario.  In particular, in the `solid'
phase at low vibration amplitude the  positional
correlation function decays slowly, consistent with an algebraic decay, and the
orientational correlation function appears to be long-ranged. The
hexagonal lattice in this regime is nearly defect free.  As the
vibration amplitude is increased, the density of defects increases
dramatically, but isolated disclinations are not observed.  In this
regime the positional correlation function decays more rapidly, while
the orientational correlation function decays very slowly, consistent with an algebraic decay. This
phase has the characteristics of the hexatic phase observed in
equilibrium 2D melting.  Finally,
at still higher amplitudes, isolated disclinations appear, and only
short ranged order is observed.  Increasing the vibration amplitude reduces the particle density in the observation region, and the density at which the transitions occur is very similar to the densities of the transitions observed in simulations of equilibrium hard-disk melting.  These results suggest that there is a deep similarity between
the coarse-grained 
dynamics of the far from equilibrium system of vibrated spheres,
and the equilibrium dynamics of hard sphere systems. 

\label{setup}
The experimental apparatus has been described
previously\cite{olafsen99,urbach01}.  Briefly, a 20 cm diameter
smooth, rigid aluminum plate is mounted horizontally and vibrated
sinusoidally in the vertical direction 
by an electromagnetic shaker.  The plate is carefully
leveled, and the acceleration is uniform across the plate to better
than 0.5\%.  The plate is machined to be flat, but in fact has a slight bowl shape, the importance of which will be discussed below.  Spheres of 316 stainless steel with a diameter of 1.588
$\pm$ 0.0032mm are placed on the plate, confined on the
sides by a 1.15cm wide aluminum rim and from above by a antistatic coated Plexiglas lid 2.54 mm
(1.6 ball diameters) above the bottom plate. The  A high resolution camera
(Pulnix TM1040) placed above the center of the plate is used to locate particle positions with a precision of less than 0.01 ball diameters (0.17 pixels) by
calculating intensity-weighted centers of the bright spots produced by reflections of the illuminating light.  The area imaged is approximately 1/8 the total area of the plate.

We have previously reported a phase diagram for this system as a
function of frequency and amplitude for different
particle densities without the lid present\cite{olafsen98,urbach01}.  The vibration amplitude is described by the dimensionless acceleration
$\Gamma=A \omega ^2 /g$, where $A$ is the amplitude of the plate
oscillation, $\omega$ is the  angular frequency of the plate oscillation, and 
$g$ is the acceleration due to gravity.   For the results described below, the vibration frequency is $\omega /2
\pi$ = 90 Hz.  We have previously shown that the phase diagram in this regime is not extremely sensitive to frequency\cite{olafsen98}.
The average two-dimensional particle
density, $\rho=N/N_{max}$, where N is the total number of particles on the plate and $N_{max}$
is the maximum number of spheres in a single hexagonal close packed
layer ($N_{max}=11270$ for this setup), is $\rho = 0.787$ for the results reported below.

\begin{figure}
\scalebox{0.6}{\includegraphics{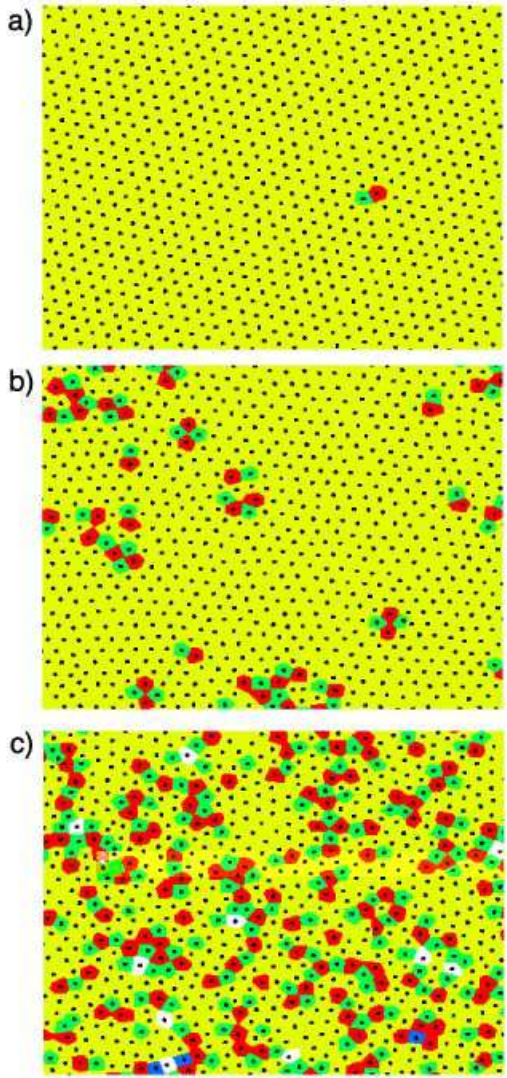}}
\caption{\label{disloc} Particle positions and topological
  defects. (a) $\Gamma=0.85$, (b) $\Gamma=1.1$, (c) $\Gamma=1.2$ 
  Neighborhoods of 6-fold coordinated particles are colored yellow, 5-fold green, 7-fold red, 8-fold white, and 4-fold blue.  Isolated disclinations are present in (c), e.g. the 7-fold coordinated particle near the top in the center or the 5-fold coordinated particle slightly left of the center.}
\end{figure}

At low accelerations, the spheres are arranged in a hexagonal lattice,
with a single domain extending across the entire cell.  Fig.
\ref{disloc}a shows the instantaneous positions of the particles at
$\Gamma=0.85$.  The particle positions fluctuate continuously, but no
particle rearrangements are observed.  In order to identify defects in
the hexagonal array, Voronoi constructions are used to identify
nearest neighbors\cite{IDL}.  The neighborhoods marked in green and red in  Fig
\ref{disloc}a represent a 5-fold and a 7-fold coordinated defect
(disclinations), which together represent a dislocation.  At this
acceleration the density of dislocations is negligible, and the
topological order is essentially  perfect.  As the amplitude of the
acceleration is increased above $\Gamma=1$, the density of topological
defects increases dramatically. Fig. \ref{disloc}b shows a snapshot
at $\Gamma=1.1$, where dislocation pairs are common and
the ordering is significantly disrupted. At still higher accelerations,
isolated disclinations are present (Fig. \ref{disloc}c), signifying the
presence of a disordered liquid.   We do not observe any evidence for phase separation, which would be expected in a discontinuous phase transition. This is in marked contrast to the behavior observed when the system can form a second layer, where a clear liquid-solid coexistence is observed\cite{prevost04,melby05}.    There is a small (less than 1\%) variation in density across the image, due to the slight bowl-shape of the plate.  The overall density in the imaged region decreases as the shaking amplitude increases, because more spheres are pushed out to the edges of the plate out of the field of view of the camera.  The decrease is approximately linear in $\Gamma$, from $\rho=0.838$ at  $\Gamma=0.85$ to $\rho=0.791$ at  $\Gamma=1.25$ 

\begin{figure}
\scalebox{0.4}{\includegraphics{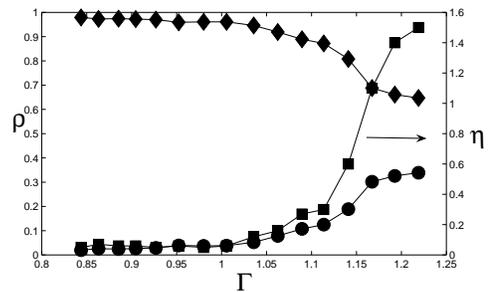}}
\caption{\label{disdensity} Left axis: Density of dislocations (circles) and
  6-fold coordinated sites (diamonds) as a function of acceleration.  Right axis:  Exponent from a fit of the orientational order correlation function to an algebraic decay, $g_6(r) \propto r^{- \eta}$ (open squares). }
\end{figure}

Figure \ref{disdensity} shows the density of dislocations (circles) and
6-fold coordinated sites (diamonds) as a function of acceleration. The
low density of dislocations for $\Gamma<1.0$ may be due to finite size
effects, and in any case has little effect on the length scales that
are accessible in our experiment.  For $\Gamma$ just over 1.0,
the density of topological defects begins to increase approximately
exponentially, doubling about every 5\% increase in $\Gamma$.
The increase slows dramatically above  $\Gamma=1.15$, which is coincident with
the appearance of isolated disclinations.

\begin{figure}[h]
{\scalebox{0.35}{\includegraphics{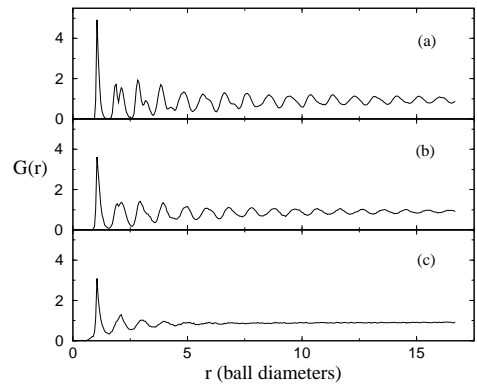}}}%
\caption{\label{position} Pair correlation function for (a) $\Gamma = 0.85$, (b) $\Gamma = 1.1$, and (c) $\Gamma = 1.1$.}
\end{figure}

The effect of the topological defects on the positional ordering of the spheres
can be seen through the pair correlation function 
$G(r) = <\rho(r') \rho(r'+r)>/<\rho(r')>^2$,
where $\rho$ is the particle density and the average is taken over the spatial variable $r'$.  Fig. \ref{position} shows the
pair correlation function for $\Gamma=0.85$, 1.1, and 1.2, the same
accelerations shown in 
Fig. \ref{disdensity}. At low accelerations, the positional
correlations decay very slowly, consistent with an algebraic decay.
As the acceleration is increased, the decay becomes considerably more
rapid, and by $\Gamma=1.2$ the positional order is clearly
short-ranged.  While unambiguous discrimination between algebraic and
exponential decays is not possible due to the the limited size of the
system, the evolution of the positional correlation function is
consistent with the KTHNY melting scenario.

\begin{figure}[h]
{\scalebox{0.35}{\includegraphics{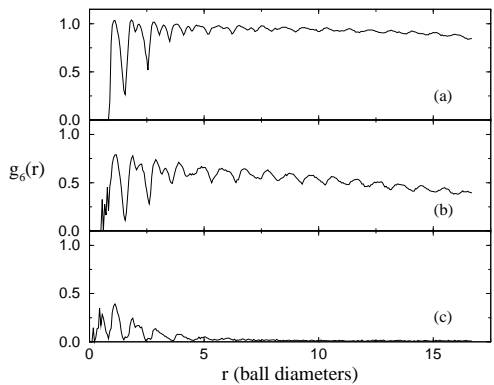}}}%
\caption{\label{angular} Orientational order correlation function  for (a) $\Gamma = 0.85$, (b) $\Gamma = 1.1$, and (c) $\Gamma = 1.1$.}
\end{figure}

The topological defects that are responsible for the decay of the
positional correlations in the hexactic phase preserve orientational
order, so the orientational correlations in the hexactic phase are
predicted to decay slowly.  Fig. \ref{angular} shows the orientational
order correlation function,
$g_6(r) = <\Psi(r') \Psi(r'+r)>/g(r)$,
where 
\begin{equation}
\Psi (\vec{r})= \sum _k ^{N} {1 \over N_j} \sum _j e^{6i \theta _{kj}}
\delta(\vec{r} - \vec{r_k})
\end{equation}
where the first  sum  runs over the $N$ particles in each image, and the
second sum runs over the $N_j$ neighbors of the $k$th particle in the
image.   $\theta _{kj}$ is the angle of the bond between the $k$th
particle and its $j$th neighbor.  At $\Gamma=0.85$, the orientational order
is essentially constant, indicating true long-range orientational
order.  At $\Gamma=1.1$, there is a noticeable decrease in the
correlation function at large separations.   As the acceleration is
increased further, the rate of the decay increases dramatically, and
by $\Gamma=1.2$, only short-range orientational order is observed,
indicating an isotropic liquid.

The correlation functions in the transition region are better fit by an 
algebraic decay than an exponential, although an exponential decay with a long correlation length cannot be ruled out.  The open squares plotted in Fig. \ref{disdensity} show the result of a fit $g_6(r) = g_6(0) r^{- \eta}$. Below $\Gamma=1$ the exponent exponent is negligible, indicating long range orientational order.  The exponent increases as the transition to an isotropic liquid is approached, as in the KTHNY melting scenario.  In the equilibrium theory, $\eta$ reaches a maximum of 1/4 at the melting transition, at which point $g_6(r)$ decays exponentially.  In our data fits producing values of $\eta$ above 1 indicate short range order (see Fig. \ref{angular}c), so within the limitations of our system size and uniformity, these results are  consistent with the KTHNY melting scenario.

The close correspondence between the melting transition we have observed and the behavior of equilibrium systems, combined  with our recent observation of a first-order-like transition to a two-layer solid\cite{prevost04,melby05},  is a strong indication that some aspects of the theory of equilibrium phase transitions should be applicable to this far-from-equilibrium system.   Both of these results are closely analogous to well understood transitions in equilibrium confined colloids, despite the presence of strong non-equilbrium effects in the velocity distributions\cite{olafsen98,olafsen99}, velocity correlations\cite{prevost02}, and energy equipartition\cite{prevost04}, although the results on the two-layer solid also show that differing dissipation rates in coexisting phases has a significant effect on the phase diagram.  The incorporation of these effects into a predictive statistical mechanics of nonequilibrium phase transitions remains an open challenge.

 The fact that the melting is driven by increasing the amplitude of the vibration, and therefore the average kinetic energy of the spheres, might seem to suggest an analogy between the granular temperature (usually defined as proportional to the mean kinetic energy per particle) and the thermodynamic temperature. However the stainless steel balls behave as hard spheres, to a very good approximation.  In equilibrium hard sphere systems, temperature is not a relevant variable, and phase transitions are driven exclusively by density changes.  In fact we believe a similar mechanism is at work in the results shown here.  As the shaking amplitude is increased, the effective density of the layer changes in two ways.  At low vibration amplitudes, the balls stay near the plate, and the horizontal distance between the centers of the spheres cannot be less than the sphere diameter.  When the shaking amplitude is increased, the balls begin to explore all of the volume available to them.  (As described above, the balls are confined from above by a lid that sits 1.6 ball diameters above the bottom plate.)  The horizontal distance between the  centers of spheres that are not in the same horizontal plane can thus be less than the ball diameter, effectively reducing the 2D area of each ball, or equivalently decreasing the normalized density.  Another, likely more important effect, arises from the slight bowl shape of the plate.  The balls tend to settle towards the center, but the amount of settling is reduced as the acceleration is increased.  Our data was acquired in the center portion of the plate, and shows a decreasing density as the amplitude is increased.  The local density at which the transition is observed and the width of the hexatic region are in fact very close to those observed in simulations of hard disk systems\cite{watanabe05}.  In an ideal system, the plate non-uniformity could be eliminated, but the ball-overlap effect would persist.  

Careful measurements of two-dimensional melting in magnetic bubble arrays have also observed a continuous hexatic-to-liquid melting transition\cite{seshadri92}.  Like the granular layer described here, the magnetic bubble arrays are far from equilibrium, in the sense that true thermal fluctuations are irrelevant.  In that system, however, each magnetic bubble is forced `microscopically', and exhibits behavior that is indistinguishable from Brownian motion.  Thus it is expected to be a good model of equilibrium behavior.  In the granular layer, however, the details of the forcing have a strong influence on the microscopic dynamics\cite{prevost02}, which typically differ significantly from those observed in equilibrium systems\cite{olafsen98,urbach01,prevost02}.
The results presented here show that despite these deviations from equilibrium systems, the `macroscopic' behavior of the granular layer follows the equilibrium melting scenario, which suggests that there may be a coarse-grained functional that could play the role of the equilibrium free energy in a quantitative theory of the observed phase transition.

We would like to thank Heather Deese for her contributions to the early stages of this project.
This work was supported by a grant from the Petroleum Research Fund, Grant DMR-9875529 from the NSF and award NNC04GA63G from NASA.

\end{document}